\documentclass[12pt,preprint]{aastex}

 \usepackage{epstopdf}
 \usepackage{caption}

 \slugcomment{Accepted Version 1}

\begin{document}

 \DeclareGraphicsExtensions{.pdf,.gif,.jpg}

 \title{Thermal Model Calibration for Minor Planets Observed with WISE/NEOWISE}
 \author{A. Mainzer\altaffilmark{1}, T. Grav\altaffilmark{2}, J. Masiero\altaffilmark{1}}
 
 \author{J. Bauer\altaffilmark{1}$^{,}$\altaffilmark{3}, E. Wright\altaffilmark{4}, R. M. Cutri\altaffilmark{3}, R. S. McMillan\altaffilmark{5}, M. Cohen\altaffilmark{6},M. Ressler\altaffilmark{1},P. Eisenhardt\altaffilmark{1}}
 \altaffiltext{1}{Jet Propulsion Laboratory, California Institute of Technology, Pasadena, CA 91109 USA}
 \altaffiltext{2}{Johns Hopkins University, Baltimore, MD}
\altaffiltext{3}{Infrared Processing and Analysis Center, California Institute of Technology, Pasadena, CA 91125, USA}
\altaffiltext{4}{UCLA Astronomy, PO Box 91547, Los Angeles, CA 90095-1547 USA}
\altaffiltext{5}{Lunar and Planetary Laboratory, University of Arizona, 1629 East University Blvd., Kuiper Space Science Bldg. \#92, Tucson, AZ 85721-0092, USA}
\altaffiltext{6}{Radio Astronomy Laboratory, 601 Campbell Hall, University of California, Berkeley, CA 94720, USA}


 \begin{abstract}
With the \emph{Wide-field Infrared Survey Explorer} \citep[WISE; ][]{Wright}, we have observed over 157,000 minor planets \citep{Mainzer11}. Included in these are a number of near-Earth objects, Main Belt Asteroids, and irregular satellites which have well-measured physical properties (via radar studies and \emph{in situ} imaging) such as diameters.  We have used these objects to validate models of thermal emission and reflected sunlight using the WISE measurements, as well as the color corrections derived in \citet{Wright} for the four WISE bandpasses as a function of effective temperature.  We have used 50 objects with diameters measured by radar or in situ imaging to characterize the systematic errors implicit in using the WISE data with a faceted spherical NEATM model to compute diameters and albedos. By using the previously measured diameters and $H$ magnitudes with a spherical NEATM model, we compute the predicted fluxes \citep[after applying the color corrections given in][]{Wright} in each of the four WISE bands and compare them to the measured magnitudes.  We find minimum systematic flux errors of 5-10\%, and hence minimum relative diameter and albedo errors of $\sim$10\% and $\sim$20\%, respectively.  Additionally, visible albedos for the objects are computed and compared to the albedos  at 3.4 $\mu$m and 4.6 $\mu$m, which contain a combination of reflected sunlight and thermal emission for most minor planets observed by WISE.  Finally, we derive a linear relationship between subsolar temperature and effective temperature, which allows the color corrections given in \citet{Wright} to be used for minor planets by computing only subsolar temperature instead of a faceted thermophysical model.  The thermal models derived in this paper are not intended to supplant previous measurements made using radar or spacecraft imaging; rather, we have used them to characterize the errors that should be expected when computing diameters and albedos of minor planets observed by WISE using a spherical NEATM model.

 \end{abstract}
 
 \section{Introduction}
The \emph{Wide-field Infrared Survey Explorer} (WISE) is a NASA Medium-class Explorer mission designed to survey the entire sky in four infrared wavelengths, 3.4, 4.6, 12 and 22 $\mu$m (denoted $W1$, $W2$, $W3$, and $W4$ respectively) \citep{Wright, Liu, Mainzer}.  The final mission data products are a multi-epoch image atlas and source catalogs that will serve as an important legacy for future research.   While WISE's primary science goals are to find the most luminous galaxies in the entire universe and to find the closest and coolest stars, the survey has yielded observations of over 157,000 minor planets, including Near-Earth Objects (NEOs), Main Belt Asteroids (MBAs), comets, Hildas, Trojans, Centaurs, and scattered disk objects \citep{Mainzer11}.  This represents an improvement of nearly two orders of magnitude more objects observed than WISE's predecessor mission, the \emph{Infrared Astronomical Satellite} \citep[IRAS;][]{Tedesco, Matson}.  The WISE survey began on 14 January, 2010, and the mission exhausted its primary tank cryogen on 5 August, 2010.  Exhaustion of the secondary tank and the start of the NEOWISE Post-Cryogenic Mission occurred on 1 October, 2010, and the survey ended on 31 January, 2011.

Infrared observations of all classes of minor planets are useful for determining size and albedo distributions, as well as thermophysical properties such as thermal inertia, the magnitude of non-gravitational forces, and surface roughness \citep[e.g.,][]{Tedesco02,Trilling,Harris09}.  Of the more than 157,000 objects detected by NEOWISE, some have physical properties such as diameter, albedo, and subsolar temperature measured by independent means such as radar observations, \emph{in situ} spacecraft visits, or stellar occultations.   We can obtain diameters and albedos for the full sample of minor planets observed by WISE, but this requires characterization of the systematic errors associated with using thermal models \citep[c.f.][]{Harris11}.  In this work, we have used a set of objects with well-known, independently measured diameters and $H$ magnitudes to test the ability of thermal models created using WISE data to accurately compute diameter and albedo.

\begin{figure}
\figurenum{1}
\includegraphics[width=3in]{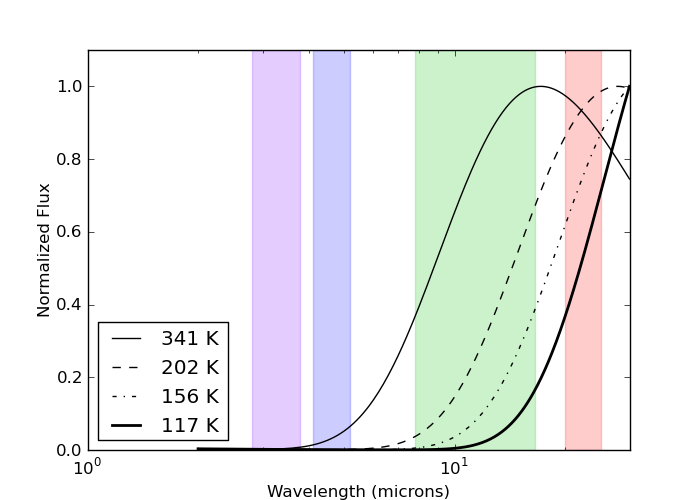}
\caption{\label{fig:bandpasses} This figure illustrates the need for the color corrections given by \citet{Wright} in order to capture the shift in zero point and effective wavelength as a function of effective temperature: the WISE bandpasses are broad, particularly $W3$. The WISE bandpasses are shown as shaded bars, and normalized fluxes are plotted for a range of different blackbody temperatures, illustrating the necessity for a color correction that varies as a function of an object's effective temperature. }
\end{figure}

Figure \ref{fig:bandpasses} shows model spectral energy distributions for minor planets with different effective temperatures with the WISE bandpasses overplotted. The color corrections given by \citet{Wright} allow the zero points to be corrected to account for the effect of flux generated by objects with varying effective temperature being observed through non-monochromatic bandpasses. The absolute zero points and isophotal wavelengths of the WISE passbands \citep{Wright,Tokunaga} are calibrated using the particular Kurucz photospheric spectrum for Vega detailed by \citet{Cohen} and validated absolutely to 1.1\% by \citet{Price}.  The width of the WISE passbands (particularly $W3$) means that it is necessary to apply a color correction for sources with a different spectrum than Vega.  For the vast majority of WISE extra-solar sources, these color correction terms are small (only a few percent for a K2V star in $W3$, for example).  However, for objects such as minor planets with effective temperatures as low as 100 K, the flux correction factor for $W3$ exceeds a magnitude \citep{Wright}, and readers are strongly encouraged to consult the WISE Explanatory Supplement \footnote{http://wise2.ipac.caltech.edu/docs/release/prelim/expsup/} and \citet{Wright} for precise values.  \citet{Wright} also find a discrepancy between red and blue calibrators in bands $W3$ and $W4$ that require a -8\% and +4\% adjustment to the zero point magnitudes in each band, respectively.  The color corrections derived in \citet{Wright} were produced by integrating over the system's relative spectral response throughput, which was measured during ground-based tests prior to launch \citep{Latvakoski}.  

We have compared the measured WISE magnitudes to magnitudes derived from spherical thermal models created for 117 objects, all of which have diameters and/or albedos measured via \emph{in situ} spacecraft measurements, radar studies, or stellar occultations.  By comparing the predicted magnitudes to the as-measured magnitudes reported by the WISE pipeline, we have verified that the color corrections derived in \citet{Wright} as a function of an object's effective temperature correctly reproduce observed physical parameters.  We have used these objects to set limits on the systematic errors for diameters and albedos derived from WISE observations of minor planets when using a spherical thermal model and the \citet{Wright} color corrections.  Finally, we have determined an empirical relationship between an object's subsolar temperature and its effective temperature, which is necessary for selecting the appropriate color correction for Solar System objects from \citet{Wright}.

 \section{Observations}
We have assembled a list of objects with well-measured diameters and $H$ magnitudes that WISE observed during the fully cryogenic portion of its mission (Table 1).   These objects were chosen because their physical measurements were obtained using methods largely independent of infrared model parameters, such as radar imaging, direct measurements from spacecraft visits or flybys, or stellar occultations.  Diameters and albedos derived from other infrared observatories such as IRAS, the \emph{Spitzer Space Telescope}, and ground-based observations were not used to verify the color corrections of \citet{Wright} due to the desire to produce an independent calibration without reference to other infrared observers' model parameter assumptions. However, for some objects, such as those observed by \citet{Shepard}, the radar diameter measurements were compared with IRAS diameters and were found to be consistent.  Of the $\sim$400 radar-observed asteroids (\emph{http://echo.jpl.nasa.gov/index.html}), a dozen or so spacecraft targets, and hundreds of occultation targets, we identified WISE observations for 117 objects.  In order to reduce the potential difficulties associated with applying a spherical thermal model to non-spherical objects, we eliminated from further consideration all objects with peak-to-peak magnitude differences larger than 0.3 magnitudes in $W3$; this reduced the sample to the 50 objects listed in Table 1. 

The observations of these objects were retrieved by querying the Minor Planet Center's (MPC) observation files to look for all instances of individual WISE detections of the desired objects that were reported using the WISE Moving Object Processing System \citep[WMOPS;][]{Mainzer11}.  The resulting set of position/time pairs were used as the basis of a query of WISE source detections in individual exposures (also known as ``Level 1b" images) using the Infrared Science Archive (IRSA).  In order to ensure that only observations of the desired moving object were returned from the query, the search radius was restricted to 0.3 arcsec from the position listed in the MPC observation file.  Additionally, since WISE collected a single exposure every 11 seconds and observes each part of the sky an average of 10 times, the modified Julian date was required to be within 2 seconds of the time specified by the MPC.  Artifacts were minimized by setting the following flags: cc\_flags = 0 or p and ph\_qual = A, B, or C (this flag indicates that the source is likely to have been a valid detection).  Objects brighter than $W3=4$ and $W4=0$ magnitudes were assumed to have flux errors equivalent to 0.2 magnitudes due to changes to the shape of the point spread function as the objects became saturated, and a linear correction was applied to the $W3$ magnitudes in this brightness regime (the WISE Explanatory Supplement contains a more detailed explanation).  Per the Explanatory Supplement, objects brighter than $W3=-2$ and $W4=-6$ were not used.  Each object had to be observed a minimum of three times in at least one WISE band, and it had to be detected at least 40\% of the time when compared to the band with the maximum number of detections (usually, though not always, $W3$). The WMOPS system is designed to reject inertially fixed objects such as stars and galaxies in bands $W3$ and $W4$. Nonetheless, the individual images at all wavelengths were compared with WISE atlas coadd and daily coadd source lists to ensure that inertially-fixed sources such as stars and galaxies were not coincident with the moving object detections.  This check is particularly important in bands $W1$ and $W2$ where the density of background objects (and hence the probability of a blended source) is higher than at longer wavelengths.  Any remaining blended sources in bands $W1$ and $W2$ were removed.  Some objects were observed at multiple epochs, and observations separated by more than three days were modeled separately.  

\section{Thermal Model and Reflected Sunlight Fits}
The thermal flux from an airless sphere is given by 
\begin{equation}F_\nu = \epsilon R^2/\Delta^2 \int_0^{\pi/2} \int_0^{2 \pi} B_\nu(T(\theta, \phi)) d\phi \sin\theta \cos\theta d\theta \end{equation}
where $\Delta$ is the object-to-observer distance, $\epsilon$ is the emissivity, $R$ is the object's radius, $B$ is the Planck function, $\theta$ is the angle from the sub-observer point to a point on the asteroid such that $\theta$ is equal to the solar phase angle $\alpha$ at the sub-solar point, and $\phi$ is an angle measured around the sub-observer point such that $\phi=0$ at the sub-solar point \citep[e.g,][]{Bhattacharya,Harris09, Delbo}.  In order to compute this flux, the temperature distribution across the body must be computed.  In the Standard Thermal Model (STM) of \citet{Lebofsky_Spencer}, the temperature of an asteroid is assumed to be maximum at the subsolar point and zero on the point opposite of this; this is the case of an object with zero thermal inertia.  In contrast, in the Fast Rotating Model (FRM) \citep{Lebofsky78,Veeder89,Lebofsky_Spencer}, the asteroid is assumed to be rotating much faster than its cooling time (i.e. high thermal inertia), resulting in a constant surface temperature across all longitudes.  The so-called ``beaming parameter" was introduced by \citet{Lebofsky} in the STM to account for the enhancement of thermal radiation observed at small phase angles.  The near-Earth asteroid thermal model (NEATM) of \citet{Harris} also uses the beaming parameter $\eta$ to account for cases intermediate between the STM and FRM models, such that \begin{equation}T(\theta,\phi) = T_{ss} [max(0,\cos\theta \cos\alpha + \sin\theta \sin\alpha \cos\phi)]^{1/4}\end{equation} where $\alpha$ the solar phase angle, and the sub-solar temperature $T_{ss}$ is given by: \begin{equation}T_{ss} = \left[\frac{S_{0}(1-A)}{\eta \epsilon \sigma}\right]^{0.25}.\end{equation}

The emissivity, $\epsilon$, is assumed to be 0.9 for all wavelengths \citep[c.f.][]{Harris09}, $A$ is the Bond albedo, $S_{0}$ is the solar flux, $\sigma$ is the Stefan-Boltzmann constant, and $\eta$ is the beaming parameter. In the STM, $\eta$ is set to 0.756 to match the occultation diameters of 1 Ceres and 2 Pallas, while in the FRM, $\eta$ is equal to $\pi$.  With NEATM, $\eta$ is a free parameter that can be fit when two or more infrared bands are available (or with only one infrared band if diameter or albedo are known \emph{a priori} as in this paper).

Each object was modeled as a set of triangular facets covering a spherical surface with diameter equal to the ground-truth measurement \citep[c.f.][]{Kaasalainen}.  Model magnitudes were computed for each WISE measurement, ensuring that the correct Sun-observer-object distances were used for each one. The temperature for each facet was computed, and the \citet{Wright} color corrections were applied to each facet.   The emitted thermal flux for each facet was calculated using NEATM along with the bandcenters and zero points given in \citet{Wright}; the temperature at the anti-subsolar point was set to 3 K, so the facets closest to this point contribute little flux.  

The objects' absolute magnitudes ($H$) were taken from \citet{Warner} and \citet{Pravec} when available; otherwise, the values were taken from the MPC's orbital element files.   Although \citet{Trilling}, \citet{Parker} and \citet{Juric} have applied a 0.3 magnitude offset to the $H$ values for objects in their respective works, we did not apply the offset to the $H$ magnitudes \citep{Spahr}.  The offset was caused by the dominance of unfiltered CCD photometry in the MPC datafiles, largely from the LINEAR survey; as more filtered measurements have become available from other surveys in recent years, the offset is no longer appropriate to use.  We will revisit these $H$ values as improved measurements become available.  

Diameters and albedos computed from the combination of diameter and $H$ from the radar, spacecraft, or occultation measurements are given in Table 1.  In addition, the Saturnian moon Phoebe had a measurement of subsolar temperature \citep{Spencer} that was also used in the thermal model.  Phoebe's $H$ value was taken from \citet{Grav}, and its $G$ value from \citet{Bauer}.  If an object had no prior independent measurement of geometric albedo ($p_{v}$) from an \emph{in situ} measurement, it was computed using the relationship \begin{equation}p_{v} = \left[\frac{1329\cdot 10^{-0.2H}}{D}\right]^{2},\end{equation} where $H$ is the object's absolute magnitude and $D$ is its diameter \citep{HarrisLagerros}.  For objects with three valid measurements in two or more WISE bands dominated by thermal emission, $\eta$ was determined using a least squares minimization.  

In general, minor planets detected by NEOWISE in bands $W1$ and $W2$ contain a mix of reflected sunlight and thermal emission.  Thus, it was necessary to incorporate an estimate of reflected sunlight into the thermal model in order to use data from bands $W1$ and $W2$.  In order to compute the fraction of reflected sunlight in bands $W1$ and $W2$, it was also necessary to compute the ratio of the infrared albedo $p_{IR}$ to the visible albedo $p_{V}$.  We make the simplifying assumption that the reflectivity is the same in both bands $W1$ and $W2$, such that $p_{IR}=p_{3.4}=p_{4.6}$;  the validity of this assumption is discussed below.  The geometric albedo $p_{V}$ is defined as the ratio of the brightness of an object observed at zero phase angle ($\alpha$) to that of a perfectly diffusing Lambertian disk of the same radius located at the same distance.  The Bond albedo ($A$) is related to the visible geometric albedo $p_{V}$ by $A\approx A_{V} = qp_{V}$, where $q$ is the phase integral and is defined such that $q=2\int \Phi(\alpha) sin(\alpha) d\alpha$. $\Phi$ is the phase curve, and $q=1$ for $\Phi=max(0,cos(\alpha))$. $G$ is the slope parameter that describes the shape of the phase curve in the $H-G$ model of \citet{Bowell} that describes the relationship between an asteroid's brightness and the solar phase angle.  For $G=0.15$, $q=0.384$.  Since $q$ is never larger than about 2/3 for any observed value of $G$, the geometric albedo $p_{V}$ is not limited to $<$1.0; it is possible to have a value as large as $\sim$1.5 without violating conservation of energy.  Conversely, values of $p_{V}$ approaching 1.0 still amount to integral reflectivity (Bond albedo) of $\sim$60\%, less than newly fallen snow, or other analogs of very high albedo.

We make the assumption that $p_{IR}$ obeys these same relationships, although it is possible that the phase integral described above varies with wavelength, so what we denote here as $p_{IR}$ for convenience may not be exactly analogous to $p_{V}$.  The flux from reflected sunlight was computed for each WISE band using the IAU phase curve correction \citep{Bowell}: \begin{equation}V(\alpha) = H + 5\log(R\Delta) - 2.5\log[(1-G)\Phi_{1}(\alpha) + G\Phi_{2}(\alpha)],\end{equation} 

where $V$ is the predicted apparent magnitude; $R$ is the heliocentric distance in AU; $\Delta$ is the geocentric distance in AU; $\alpha$ is the Sun-observer-object angle; $G$ is slope parameter for the magnitude-phase relationship; and \begin{equation}\Phi_{i}(\alpha) = \exp[-A_{i}\tan(\frac{1}{2}\alpha)^{B_{i}}],\end{equation} where \begin{equation}i = 1,2, A_{1} = 3.33, B_{1} = 0.63, A_{2} = 1.87, B_{2} = 1.22.\end{equation}

Unless a direct measurement of $G$ was available from \citet{Warner} or \citet{Parker}, we assumed a $G$ value of 0.15.  Finally, the weighted averages of the model magnitude, measured WISE magnitude and $T_{ss}$ were computed for each object by grouping together observations no more than a three day gap between them.  This was done to ensure that NEOs, which can have significant changes in distance over short times, were modeled accurately.

Error bars on the model magnitudes and subsolar temperatures were determined for each object by running 100 Monte Carlo trials that varied the objects' $H$ values by the error bars given in Table 1, their diameters by the error bars specified in Table 1, and the WISE magnitudes by their error bars using Gaussian probability distributions.  The minimum magnitude error for all WISE measurements fainter than $W3=4$ and $W4=3$ magnitudes was 0.03 magnitudes, per the in-band repeatability measured in \citet{Wright}. For objects brighter than $W3=4$ and $W4=0$, the error bars were increased to 0.2 magnitudes, as these magnitudes represent the onset of saturation.  Additionally, a linear correction was applied to objects with $-2 < W3 < 4$. The effect is thought to be caused by changes in the point spread function observed when images start to enter saturation (see the WISE Explanatory Supplement for a discussion of the effects of saturation on photometry).  We have set the error bars to 0.2 magnitudes right at these limits in order to conservatively capture residual errors not fixed by the simple linear correction applied.  Objects with $W3<-2$ and $W4<-6$ were too bright to be used.  The error bar for each object's model magnitude was equal to the standard deviation of all the Monte Carlo trial values.  If a measurement of $H$ was not available from \citet{Warner} or \citet{Pravec}, the error in $H$ was taken to be 0.3 magnitudes \citep{Spahr}.  

\begin{figure}
\figurenum{2}
\includegraphics[width=6in]{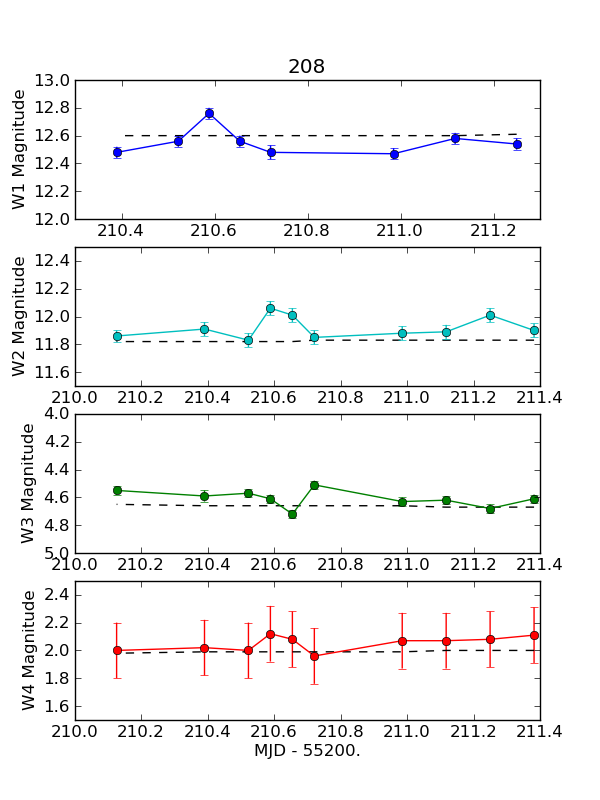}
\caption{\label{fig:lightcurve}WISE observations as a function of modified Julian date as well as a spherical NEATM model fit are shown for a typical calibrator object, (208) Lacrimosa.  NEOWISE detections of this asteroid span $\sim$17 hours.  The $W1$ points are shown in dark blue; $W2$ in cyan, $W3$ in green, and $W4$ in red. }
\end{figure}

\begin{figure}
\figurenum{3abcd}
\includegraphics[width=6in]{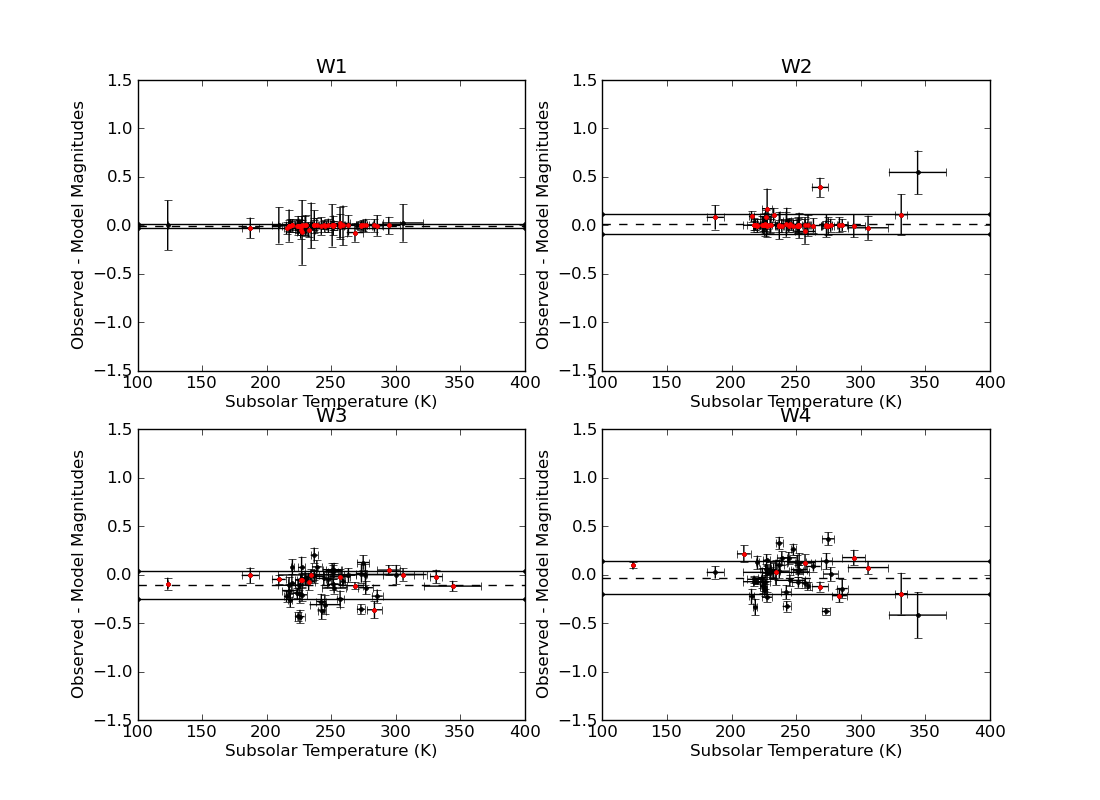}
\caption{\label{fig:calibrators}With WISE, we have observed objects with diameters that have been measured either with radar observations, \emph{in situ} spacecraft visits, or by stellar occultations (see Table 1).  After applying the color corrections specified in \citet{Wright}, we can compare the observed WISE magnitudes to the fluxes predicted by a thermophysical model for bands $W1$ through $W4$ (Figures (a)-(d), respectively). The dashed line shows the weighted mean value of all the points; the dotted lines are the 1-$\sigma$ errors.  Objects with WISE measurement errors that are less than 0.1 magnitudes are shown as red dots; objects with errors greater than 0.1 magnitudes are shown as black dots.}
\end{figure}

Figure \ref{fig:lightcurve} shows an example of an asteroid's WISE magnitudes as well as the resulting thermal model fit.   Figures \ref{fig:calibrators} show the difference between the average measured WISE magnitudes and the model magnitudes for bands $W1$ through $W4$ using the color corrections given in \citet{Wright}.  In addition, we adjusted the $W3$ effective wavelength blueward by 4\% from 11.5608$\mu$m to 11.0984 $\mu$m, the $W4$ effective wavelength redward by 2.5\% from 22.0883 $\mu$m to 22.6405 $\mu$m, and we included the -8\% and +4\% offsets to the $W3$ and $W4$ magnitude zeropoints (respectively) due to the red-blue calibrator discrepancy reported by \citet{Wright}. The weighted means of the differences between observed and model magnitudes shown in Figure \ref{fig:calibrators} are given in Table 2, and they are near zero for most objects.  The apparent trend below zero for the objects with $T_{ss}$ larger than $\sim$300 K could be due to any of the following causes: 1) imperfect knowledge of the system relative spectral response curves, particularly in $W4$; 2) inaccuracies in either $H$ values or diameters for some of the objects; 3) the temperature distribution may not precisely follow that given in Equation 2.  This is less important for objects observed at low phase angles.  The warmer objects observed by WISE tend to be NEOs observed at higher phase angles, which could lead to the subsolar point contributing significantly more flux in the shorter wavelengths and the nightside of the object more flux at longer wavelengths.  Nevertheless, most of the predicted magnitudes are in good agreement with the observed magnitudes, indicating that the procedure given in \citet{Wright} for color correction is adequate. 

\begin{figure}
\figurenum{4}
\includegraphics[width=3in]{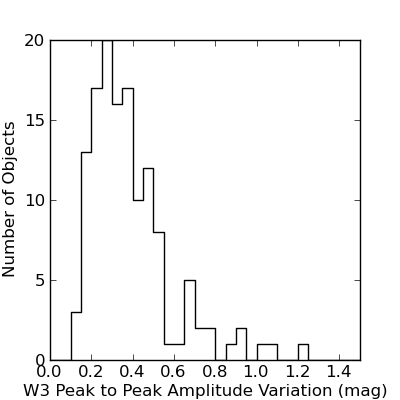}
\caption{\label{fig:lightcurve_amplitudes}The distribution of peak-to-peak amplitudes in $W3$ for 117 objects with independently measured diameters peaks at $\sim$0.25 magnitudes.  Objects with peak-to-peak amplitudes larger than 0.3 magnitudes were excluded from our analysis.}
\end{figure}

Although many of the calibrator objects are known to be non-spherical (e.g. from radar shape modeling or optical lightcurves), the WISE observations generally consisted of $\sim$10-12 observations per object uniformly distributed over $\sim$36 hours \citep{Wright, Mainzer}, so on average, a wide range of rotational phases was sampled.  Although the variation in effective spherical diameter resulting from rotational effects tends to be averaged out, caution must be exercised when interpreting effective diameter results using spherical models for objects that are known to have large-amplitude lightcurve variations. Figure \ref{fig:lightcurve_amplitudes} shows the distribution of peak-to-peak $W3$ amplitudes for the entire sample of 117 objects.  We have compared the mean differences between observed and model magnitudes for the entire sample of 117 objects to the sample of only those 50 objects with peak-to-peak $W3$ amplitudes $<0.3$ magnitudes, and there is no significant difference in the result given in Table 2.  Nevertheless, to avoid any potential difficulties associated with applying spherical models to elongated objects, we excluded objects with peak-to-peak $W3$ amplitudes from further analysis. (Figure \ref{fig:calibrators} shows the difference between observed and model magnitudes for only the 50 low-amplitude objects).  For the objects in Table 1 for which pole orientations could be found in the lightcurve database of \citet{Warner}, we computed the visible lightcurve amplitude at the aspect observed by WISE and found that for these objects, the apparent amplitudes were $<\sim$0.3 magnitudes.

\section{Systematic Diameter and Albedo Errors}
The offsets and errors given in Table 2 can be regarded as the minimum systematic errors in magnitude for minor planets observed by WISE/NEOWISE.  Since diameter is proportional to the square root of the thermal flux (Equation 1), the minimum systematic diameter error due to uncertainties in the color correction is proportional to one-half the error in flux.  These magnitude errors result in a minimum systematic error of $\sim$5-10\% for diameters derived from WISE data; they are of similar magnitude to the diameter uncertainties of most of the underlying radar and spacecraft measurements, which are $\sim$10\% (references are given in Table 1).  Albedo is proportional to $D^{2}$ (Equation 4), and so it is linearly proportional to flux to first order.  Therefore, minimum systematic errors on albedos computed from WISE observations are $\sim$10-20\%, subject to the assumption that spherical effective diameters can be computed for non-spherical shapes (future work will model the objects as non-spherical shapes and will use shape models and rotational information produced by lightcurve and radar studies).  These should be regarded as minimum errors in cases of good signal-to-noise detections when the beaming parameter and the infrared albedo can be fit.  It should also be noted that these error estimates apply only to objects as distant as Saturn, as the most distant object we have considered herein is Saturn's moon Phoebe.  Objects observed by WISE at greater distances (and therefore lower temperatures) may be subject to additional errors.  

\section{Converting Subsolar Temperature to Effective Temperature}
The color corrections specified in \citet{Wright} are given as a function of input spectra that are assumed to blackbodies of different effective temperatures ($T_{eff}$).  In order to use the \citet{Wright} color corrections by computing only the subsolar temperature rather than a faceted thermophysical model for each object, we have computed the relationship between $T_{ss}$ and ($T_{eff}$).  The total flux ($F(W_{n})$, where $n=1,...,4$) was computed using NEATM by applying a color correction to each facet on the sphere based on the facet's blackbody temperature.  The equivalent flux at the center wavelength of each bandpass is computed, $F(\lambda_{isophot})$; since this is a monochromatic flux, the color corrections are unity.  We then derive an effective flux correction ($f_{eff}$) which is given by \begin{equation}f_{eff} = F(W_{n}) / F(\lambda_{isophot}).\end{equation}  The final step is to find the $T_{eff}$ that gives the blackbody flux correction $f(T_{eff}) = f_{eff}$.

\begin{figure}
\figurenum{5}
\includegraphics[width=3in]{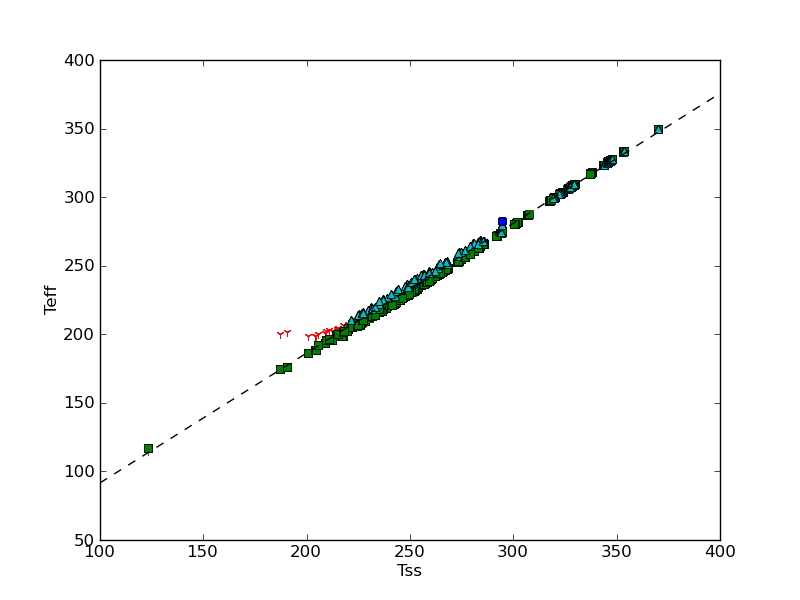}
\caption{\label{fig:Tss_vs_Teff}The correlation between subsolar temperature and effective temperature is well-described by a linear relationship. $W1 - W4$ detections are shown in blue circles; $W2$ as cyan triangles; $W3$ as green squares; and $W4$ as red inverted triangles. Each object can contribute up to four points on the plot, depending on the number of bands in which it was detected.}
\end{figure}

Figure \ref{fig:Tss_vs_Teff} shows the relationship between $T_{ss}$ and $T_{eff}$; a least-squares fit to these points yields the relationship \begin{equation} T_{eff}(K) = 0.95 T_{ss}(K) - 3.01. \end{equation}  We excluded objects with more than 20\% reflected sunlight in a given band because their color corrections will be dominated by the small corrections needed for a G star \citep{Wright}.  A single object can contribute up to four points on this plot, one for each band it is detected in with less than 20\% reflected sunlight.  Equation 9 provides a shortcut method to flux correct the WISE magnitudes in lieu of performing the facet-by-facet correction described above with negligible additional error.

\section{Albedo as a Function of Wavelength}
\begin{figure}
\figurenum{6}
\includegraphics[width=3in]{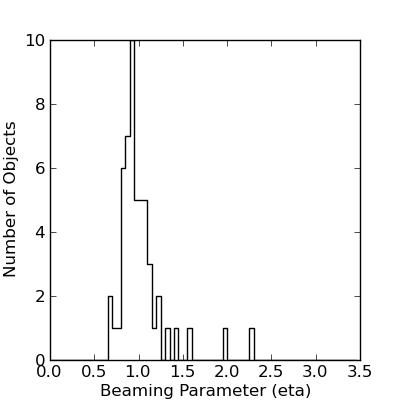}
\caption{\label{fig:beaming_hist}The beaming parameter $\eta$ values resulting from our NEATM fits are compared with solar phase angle $\alpha$ for the 50 objects listed in Table 1. The mean value for $\eta=0.96 \pm 0.28$. }
\end{figure}

\begin{figure}
\figurenum{7}
\includegraphics[width=3in]{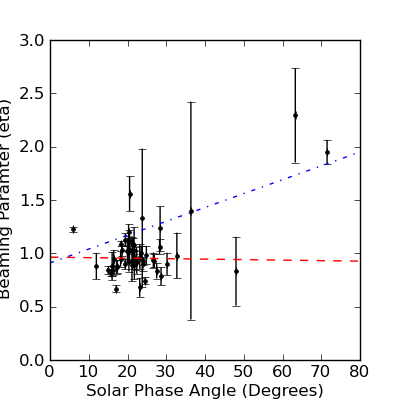}
\caption{\label{fig:phase_v_eta} The beaming parameter vs. phase angle.  The best linear fit to our data is $\eta = -0.00054 \alpha + 0.97$ and is plotted as a red dashed line.  The relationship from \citet{Wolters} is given by $\eta = 0.013\alpha + 0.91$ and is shown as a blue dash-dot line.  Future work will examine the relationship between $\eta$ and $\alpha$ using the full WISE dataset, which includes many more objects over a wide range of phase angles. }
\end{figure}

For objects that were detected according to the criteria described above in either $W1$ or $W2$, we computed the albedo at these wavelengths ($p_{IR}$, assuming $p_{3.4 \mu m} = p_{4.6 \mu m}$) in addition to $p_{V}$.  Although \citet{Trilling} and \citet{Harris09} assume that the albedo at 3.4 $\mu$m is 1.4 times higher than $p_{V}$, this result is based on seven S-type Main Belt asteroids observed at 3 $\mu$m by \citet{Rivkin}.  \citet{Harris11} use observations of a number of NEOs observed by \emph{Spitzer} and find $p_{IR} / p_{V}$ consistent with 1.4, using the relationship between $\eta$ and $\alpha$ defined by \citet{Wolters}.  We have not used the \citet{Wolters} relationship to derive $\eta$ and instead have allowed it to vary.  Since we know the diameter and can derive $p_{V}$ from Equation 4, we fit $\eta$ independently for each object in Table 1.  Figure \ref{fig:beaming_hist} shows the histogram of beaming parameters, and Figure \ref{fig:phase_v_eta} shows $\eta$ as a function of solar phase angle for the objects described in Table 1; however, we expect to significantly improve upon this result in a future work by using the general population of asteroids observed by WISE over a wide range of phase angles.  

\begin{figure}
\figurenum{8}
\includegraphics[width=3in]{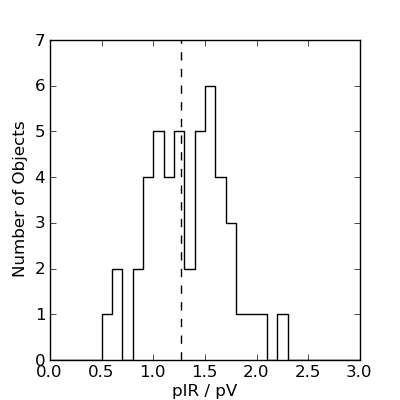}
\caption{\label{fig:pVpIR}The ratio of the albedo at $W1$ and $W2$ (we assume the albedo is the same in both of these bands) compared to the albedo at visible wavelengths as a function of subsolar temperature.  The dashed line indicates the weighted mean value of $p_{IR} / p_{V}=1.27\pm0.37$.}
\end{figure}

Figure \ref{fig:pVpIR} shows $p_{IR} / p_{V}$ for the objects detected in $W1$ and $W2$; as shown in Table 1, they represent a mix of NEOs, MBAs, and irregular satellites.  The weighted average of the ratio of $p_{IR} / p_{V}$ is $1.27  \pm 0.37$.  In computing $p_{IR} / p_{V}$, we have made the simplifying assumption that $G$ does not vary with wavelength, although it is known that $G$ increases slightly, from 0.15 to 0.20, when going from $B$ to $R$ bands \citep{BowellLumme}. However, in some cases, the variation in albedo with wavelength could be due to material and/or chemical properties \citep[c.f. Phoebe;][]{Clark}. A future work comparing both the infrared and visible albedos with known taxonomic classifications is in progress.  

\section{Conclusions}
The calibration methodology described in this work is not unique to the WISE infrared data; however, the uniquely large set of minor planet observations afforded by WISE/NEOWISE has permitted characterization of the systematic errors produced when applying spherical NEATM models to a number of objects with previously measured diameters that span a wide range of populations. We have studied NEOs, Main Belt asteroids, and irregular satellites.  In particular, the selection of an effective temperature to use when applying a literature color correction can cause large changes in the resultant best-fit diameter unless the relation between $T_{ss}$ and $T_{eff}$ is well-understood or a complete faceted model is employed. 

The color corrections described in \citet{Wright} have been used to produce an estimate of the minimum systematic errors in magnitudes for minor planets detected by WISE/NEOWISE \citep{Mainzer11}.  We have shown that the minimum diameter errors resulting from residual uncertainties in the color corrections and assumptions regarding $G$, $H$, and albedo are comparable to the uncertainties in the diameters measured by radar or \emph{in situ} spacecraft imaging (i.e. $\sim$10\%); albedos can be determined to $\sim$20\% when good signal-to-noise multi-band WISE measurements and visible data are available.  However, we note that objects more distant than the Trojan asteroids may be subject to increased systematic errors, as the most distant object studied in this work was Saturn's moon Phoebe.   Our model includes an estimate of reflected sunlight computed in all four bands, and we have computed albedos at 3.4 and 4.6 $\mu$m as well as at visible wavelengths, which should yield interesting compositional insights when compared with spectroscopic taxonomic data.  Areas for future improvement of thermal models include studying the effects of observing at high phase angles, modeling non-spherical shapes, and allowing both $G$ and albedo to vary as a function of wavelength.   All observations were processed with the Pass 1 (Preliminary) version of the WISE data processing pipeline; as the final version of the pipeline becomes available, which incorporates many improvements to instrumental calibration, we will revisit the thermal models for these objects.  We have also derived a simple linear relationship between subsolar temperature and the effective temperature used in \citet{Wright}, which should facilitate appropriate choices of color corrections for the WISE bandpasses.  The thermal models computed herein are not intended to supplant the diameters measured by radar, in situ imaging or occultations, but rather they provide insight into the ability of spherical NEATM thermal models to accurately determine diameter and albedo when applied to the general population of WISE-observed minor planets for which these parameters have not been previously determined.  This demonstrates that the WISE dataset offers a powerful new means of characterizing physical parameters of minor planets with great accuracy.

\section{Acknowledgments}

\acknowledgments{We thank L. Benner, M. Busch and M. Shepard for useful discussions and for providing diameter information on a number of objects in advance of publication.  This publication makes use of data products from the \emph{Wide-field Infrared Survey Explorer}, which is a joint project of the University of California, Los Angeles, and the Jet Propulsion Laboratory/California Institute of Technology, funded by the National Aeronautics and Space Administration.  This publication also makes use of data products from NEOWISE, which is a project of the Jet Propulsion Laboratory/California Institute of Technology, funded by the Planetary Science Division of the National Aeronautics and Space Administration. We thank our referee, A. W. Harris of Pasadena, California, for constructive comments that materially improved this work.  We gratefully acknowledge the extraordinary services specific to NEOWISE contributed by the International Astronomical Union's Minor Planet Center, operated by the Harvard-Smithsonian Center for Astrophysics, and the Central Bureau for Astronomical Telegrams, operated by Harvard University.  We also thank the worldwide community of dedicated amateur and professional astronomers devoted to minor planet follow-up observations. This research has made use of the NASA/IPAC Infrared Science Archive, which is operated by the Jet Propulsion Laboratory, California Institute of Technology, under contract with the National Aeronautics and Space Administration. This research has made use of NASA's Astrophysics Data System.}

 \clearpage

\begin{deluxetable}{llllllll}
\tabletypesize{\small}
\tablecolumns{8}
\tablecaption{Spherical NEATM models were created for 50 objects ranging from NEOs to irregular satellites in order to characterize the accuracy of diameter and albedo errors derived from NEOWISE data.  The diameters and $H$ values used to fit each object from the respective source data (either radar, spacecraft imaging, or occultation) are given.  Objects that are listed twice were observed at two epochs separated by more than three days.  When observations were separated by more than three days, a separate calculation was made.}
\tablehead{
\colhead{Object} & \colhead{$D$ (km)} &\colhead{H}  & \colhead{$p_{v}$} &\colhead{$p_{IR}$} &\colhead{$T_{ss}(K)$} & \colhead{$\eta$} & \colhead{Ref}
}
\startdata
5 & $ 115\pm   12 $ &   6.9 & $  0.25\pm  0.05 $ & $  0.33\pm  0.07 $ & $ 232.0\pm   3.0 $ & $  0.99\pm  0.07 $ & \tablenotemark{a}\\
8 & $ 140\pm  14 $ &   6.4 & $  0.26\pm  0.05 $ & $  0.44\pm  0.06 $ & $ 285.0\pm   4.8 $ & $  0.79\pm  0.03 $ & \tablenotemark{a}\\
13 & $ 227\pm  38 $ &   6.7 & $  0.07\pm  0.01 $ & $  0.04\pm  0.18 $ & $ 252.2\pm  13.6 $ & $  0.89\pm  0.21 $ & \tablenotemark{c}\\
22 & $ 143\pm  14 $ &   6.5 & $  0.17\pm  0.06 $ & $  0.29\pm  0.06 $ & $ 226.0\pm   3.2 $ & $  1.11\pm  0.09 $ & \tablenotemark{a}\\
27 & $ 118\pm  17 $ &   7.0 & $  0.20\pm  0.03 $ & $  0.46\pm  0.24 $ & $ 236.9\pm   5.5 $ & $  1.08\pm  0.10 $ & \tablenotemark{c}\\
31 & $ 280\pm  23 $ &   6.7 & $  0.05\pm  0.01 $ & $  0.08\pm  0.01 $ & $ 217.3\pm   2.6 $ & $  0.88\pm  0.05 $ & \tablenotemark{c}\\
36 & $ 103\pm   1 $ &   8.5 & $  0.07\pm  0.01 $ & $  0.06\pm  0.01 $ & $ 283.5\pm   0.8 $ & $  0.83\pm  0.01 $ & \tablenotemark{c}\\
39 & $ 163\pm   16 $ &   6.1 & $  0.25\pm  0.06 $ & $  0.48\pm  0.10 $ & $ 260.8\pm   4.2 $ & $  0.83\pm  0.07 $ & \tablenotemark{a}\\
47 & $ 138\pm  13 $ &   7.8 & $  0.07\pm  0.02 $ & $  0.06\pm  0.01 $ & $ 227.0\pm   3.2 $ & $  1.13\pm  0.07 $ & \tablenotemark{d}\\
53 & $ 115\pm   8 $ &   8.8 & $  0.04\pm  0.00 $ & $  0.04\pm  0.01 $ & $ 274.7\pm   2.9 $ & $  1.06\pm  0.05 $ & \tablenotemark{c}\\
54 & $ 142\pm  14 $ &   7.7 & $  0.05\pm  0.01 $ & $  0.07\pm  0.01 $ & $ 280.9\pm   5.5 $ & $  0.84\pm  0.07 $ & \tablenotemark{a}\\
83 & $  84\pm   8 $ &   8.7 & $  0.09\pm  0.04 $ & $  0.13\pm  0.03 $ & $ 259.1\pm   4.5 $ & $  0.92\pm  0.07 $ & \tablenotemark{c}\\
85 & $ 163\pm  16 $ &   7.6 & $  0.06\pm  0.02 $ & $  0.07\pm  0.04 $ & $ 224.8\pm   3.8 $ & $  1.03\pm  0.07 $ & \tablenotemark{a}\\
94 & $ 187.5\pm  27 $ &   7.6 & $  0.05\pm  0.02 $ & $  0.06\pm  0.02 $ & $ 219.9\pm   4.9 $ & $  1.09\pm  0.10 $ & \tablenotemark{d}\\
97 & $  83\pm   6 $ &   7.7 & $  0.21\pm  0.04 $ & $  0.29\pm  0.02 $ & $ 273.0\pm   3.4 $ & $  0.74\pm  0.05 $ & \tablenotemark{b}\\
105 & $ 119\pm  11 $ &   8.6 & $  0.05\pm  0.02 $ & $  0.03\pm  0.01 $ & $ 294.5\pm   5.1 $ & $  0.90\pm  0.07 $ & \tablenotemark{b}\\
114 & $ 100\pm   9 $ &   8.3 & $  0.09\pm  0.02 $ & $  0.14\pm  0.04 $ & $ 263.3\pm   4.5 $ & $  0.98\pm  0.07 $ & \tablenotemark{c}\\
114 & $ 100\pm  16 $ &   8.3 & $  0.09\pm  0.03 $ & $  0.13\pm  0.05 $ & $ 251.6\pm   7.0 $ & $  1.01\pm  0.12 $ & \tablenotemark{c}\\
128 & $ 188\pm  29 $ &   7.5 & $  0.05\pm  0.03 $ & $  0.05\pm  0.02 $ & $ 247.8\pm   6.3 $ & $  0.96\pm  0.10 $ & \tablenotemark{c}\\
135 & $  77\pm   8 $ &   8.2 & $  0.15\pm  0.06 $ & $  0.25\pm  0.05 $ & $ 223.7\pm   2.8 $ & $  1.20\pm  0.08 $ & \tablenotemark{e}\\
139 & $ 164\pm  19 $ &   7.9 & $  0.04\pm  0.01 $ & $  0.06\pm  0.01 $ & $ 244.2\pm   4.3 $ & $  0.93\pm  0.07 $ & \tablenotemark{d}\\
145 & $ 151\pm  23 $ &   8.1 & $  0.04\pm  0.01 $ & $  0.04\pm  0.01 $ & $ 236.8\pm   5.3 $ & $  1.09\pm  0.11 $ & \tablenotemark{d}\\
194 & $ 169\pm  18 $ &   7.7 & $  0.05\pm  0.02 $ & $  0.06\pm  0.02 $ & $ 237.0\pm   4.4 $ & $  0.90\pm  0.07 $ & \tablenotemark{c}\\
198 & $  57\pm   7 $ &   8.3 & $  0.26\pm  0.06 $ & $  0.37\pm  0.07 $ & $ 245.0\pm   5.7 $ & $  0.89\pm  0.11 $ & \tablenotemark{d}\\
208 & $  45\pm   5 $ &   9.3 & $  0.17\pm  0.06 $ & $  0.29\pm  0.06 $ & $ 224.6\pm  7.4 $ & $  1.16\pm  0.18 $ & \tablenotemark{a}\\
208 & $  45\pm   5 $ &   9.3 & $  0.16\pm  0.05 $ & $  0.29\pm  0.06 $ & $ 231.1\pm   6.7 $ & $  1.06\pm  0.14 $ & \tablenotemark{a}\\
211 & $ 143\pm  13 $ &   7.9 & $  0.06\pm  0.02 $ & $  0.07\pm  0.01 $ & $ 217.4\pm   3.0 $ & $  0.94\pm  0.06 $ & \tablenotemark{d}\\
230 & $ 109\pm  16 $ &   7.3 & $  0.17\pm  0.02 $ & $  0.30\pm  0.11 $ & $ 251.7\pm   6.5 $ & $  0.97\pm  0.10 $ & \tablenotemark{c}\\
266 & $ 109\pm   7 $ &   8.5 & $  0.06\pm  0.01 $ & $  0.05\pm  0.01 $ & $ 252.2\pm   2.1 $ & $  0.92\pm  0.04 $ & \tablenotemark{d}\\
308 & $ 144\pm  13 $ &   8.2 & $  0.05\pm  0.01 $ & $  0.08\pm  0.02 $ & $ 238.8\pm   3.1 $ & $  1.08\pm  0.06 $ & \tablenotemark{e}\\
345 & $  99.\pm   9 $ &   8.7 & $  0.06\pm  0.01 $ & $  0.06\pm  0.01 $ & $ 276.7\pm   4.4 $ & $  0.93\pm  0.06 $ & \tablenotemark{e}\\
350 & $  99.5\pm   5 $ &   8.4 & $  0.08\pm  0.02 $ & $  0.08\pm  0.01 $ & $ 227.4\pm   1.5 $ & $  0.87\pm  0.03 $ & \tablenotemark{e}\\
444 & $ 163\pm  36 $ &   7.8 & $  0.05\pm  0.04 $ & $  0.06\pm  0.06 $ & $ 229.7\pm  10.2 $ & $  0.95\pm  0.15 $ & \tablenotemark{d}\\
488 & $ 150\pm  21 $ &   7.8 & $  0.06\pm  0.01 $ & $  0.06\pm  0.02 $ & $ 219.5\pm   4.4 $ & $  0.83\pm  0.07 $ & \tablenotemark{d}\\
522 & $  84\pm   9 $ &   9.1 & $  0.06\pm  0.03 $ & $  0.09\pm  0.02 $ & $ 215.4\pm   4.5 $ & $  0.84\pm  0.07 $ & \tablenotemark{e}\\
566 & $ 134\pm  15 $ &   8.0 & $  0.06\pm  0.02 $ & $  0.09\pm  0.02 $ & $ 218.3\pm   3.6 $ & $  0.81\pm  0.06 $ & \tablenotemark{e}\\
654 & $ 127\pm  13 $ &   8.5 & $  0.04\pm  0.01 $ & $  0.05\pm  0.01 $ & $ 250.7\pm   3.9 $ & $  0.95\pm  0.07 $ & \tablenotemark{d}\\
704 & $ 312\pm  17 $ &   5.9 & $  0.08\pm  0.02 $ & $  0.09\pm  0.01 $ & $ 225.8\pm   2.0 $ & $  0.88\pm  0.03 $ & \tablenotemark{d}\\
704 & $ 312\pm  30 $ &   5.9 & $  0.08\pm  0.03 $ & $  0.10\pm  0.01 $ & $ 224.4\pm   2.9 $ & $  0.84\pm  0.05 $ & \tablenotemark{d}\\
925 & $  58\pm   6 $ &   8.3 & $  0.25\pm  0.05 $ & $  0.38\pm  0.10 $ & $ 244.6\pm   4.5 $ & $  0.90\pm  0.08 $ & \tablenotemark{a}\\
951 & $  12\pm   1 $ &  11.5 & $  0.33\pm  0.04 $ & $  0.47\pm  0.07 $ & $ 268.4\pm   4.6 $ & $  0.68\pm  0.05 $ & \tablenotemark{g}\\
1512 & $  65\pm   7 $ &   9.6 & $  0.06\pm  0.03 $ & $  0.00\pm  0.00 $ & $ 243.4\pm   4.7 $ & $  0.66\pm  0.06 $ & \tablenotemark{e}\\
1627 & $   9\pm   1 $ &  12.9 & $  0.15\pm  0.03 $ & $  0.22\pm  0.01 $ & $ 256.8\pm   4.4 $ & $  1.23\pm  0.10 $ & \tablenotemark{h}\\
1866 & $   8.7\pm   1 $ &  12.7 & $  0.19\pm  0.07 $ & $  0.30\pm  0.05 $ & $ 209.4\pm   7.8 $ & $  1.56\pm  0.25 $ & \tablenotemark{i}\\
2867 & $   5.3\pm   1.2 $ &  13.4 & $  0.28\pm  0.13 $ & $  0.56\pm  0.29 $ & $ 234.5\pm  23.8 $ & $  1.33\pm  0.55 $ & \tablenotemark{j}\\
7335 & $   0.9\pm   0.4 $ &  17.0 & $  0.31\pm  0.30 $ & $  0.40\pm  0.30 $ & $ 276.4\pm  43.0 $ & $  1.40\pm  0.98 $ & \tablenotemark{k}\\
68216 & $   1.4\pm   0.2 $ &  16.3 & $  0.27\pm  0.10 $ &   --   & $ 300.6\pm  13.4 $ & $  1.02\pm  0.22 $ & \tablenotemark{l}\\
68216 & $   1.4\pm   0.2 $ &  16.3 & $  0.31\pm  0.19 $ &   --  & $ 305.0\pm  21.7 $ & $  2.24\pm  0.55 $ & \tablenotemark{l}\\
164121 & $   1.1\pm   0.3 $ &  16.2 & $  0.36\pm  0.20 $ & $  0.50\pm  0.26 $ & $ 344.1\pm  43.2 $ & $  0.88\pm  0.39 $ & \tablenotemark{m}\\
Himalia & $ 136.0\pm  12 $ &   7.9 & $  0.07\pm  0.01 $ & $  0.07\pm  0.01 $ & $ 187.3\pm   4.5 $ & $  0.88\pm  0.10 $ & \tablenotemark{n}\\
2005 CR37 & $   1.0\pm   0.2 $ &  18.9 & $  0.05\pm  0.02 $ & $  0.06\pm  0.03 $ & $ 300.4\pm  26.1 $ & $  0.98\pm  0.31 $ & \tablenotemark{o}\\
2008 EV5 & $   0.4\pm   0.0 $ &  19.7 & $  0.13\pm  0.01 $ & $  0.18\pm  0.01 $ & $ 331.1\pm   6.8 $ & $  1.96\pm  0.15 $ & \tablenotemark{p}\\
Phoebe & $ 213.2\pm   1.3 $ &   6.6 & $  0.09\pm  0.02 $ & $  0.06\pm  0.00 $ & $ 123.4\pm   0.3 $ & $  1.23\pm  0.01 $ & \tablenotemark{q}\\
\enddata

\tablenotetext{a}{\citep{Durech}}
\tablenotetext{b}{\citep{Magri}}
\tablenotetext{c}{\citep{Magri07}}
\tablenotetext{d}{\citep{Shevchenko}}
\tablenotetext{e}{\citep{Timerson}}
\tablenotetext{f}{\citep{Shepard}}
\tablenotetext{g}{\citep{Thomas}}
\tablenotetext{h}{\citep{Ostro90}}
\tablenotetext{i}{\citep{Ostro91}}
\tablenotetext{j}{\citep{Keller}}
\tablenotetext{k}{\citep{Mahapatra}}
\tablenotetext{l}{\citep{Benner}}
\tablenotetext{m}{\citep{Benner08}}
\tablenotetext{n}{\citep{PorcoHimalia}}
\tablenotetext{o}{\citep{Benner06}}
\tablenotetext{p}{\citep{Busch11}}
\tablenotetext{q}{\citep{PorcoPhoebe,Clark}}
\end{deluxetable}

\begin{deluxetable}{cccc}
\tabletypesize{\small}
\tablecolumns{4}
\tablecaption{The mean offset ($\Delta$m) and standard deviation ($\sigma$m) of the observed - model magnitudes for the calibrator objects shown in Figure \ref{fig:calibrators}. N is the number of observations used.}
\tablehead{
\colhead{Band} & {N} & {$\Delta$m} & {$\sigma$m}
}
\startdata
W1 &          46 & 0.00 &  0.02 \\
W2 &          47 &  0.01 &  0.10 \\
W3 &          52 & -0.11 &  0.14 \\
W4 &          50 & -0.03 &  0.17 \\
\enddata
\end{deluxetable}

\end{document}